\documentclass[aps,pre,twocolumn,superscriptaddress,showkeys,showpacs]{revtex4-2}
\usepackage[utf8]{inputenc}
\usepackage{bm}
\usepackage{amssymb}
\usepackage{mathtools}
\usepackage{amsmath}
\usepackage{xcolor}
\usepackage{enumitem}
\usepackage{natbib}
\usepackage{hyperref}
\usepackage{lipsum}
\usepackage{appendix}

\begin{document}
\title{Steering the dynamics by controlling the temporal interaction network}


\author{Melvyn Tyloo}
\affiliation{Living Systems Institute, University of Exeter, Exeter, United Kingdom
and Department of Mathematics and Statistics, Faculty of Environment, Science, and Economy, University of Exeter,
Exeter, United Kingdom}

\date{\today}
\begin{abstract}
Many real-world coupled dynamical systems have their interaction structure and strength that evolve or adapt over time. 
Here, we investigate how one can control the state of a system by tuning its temporal interaction network. 
We present a framework based on nonlinear optimal control, where one has control over the time-dependent coupling matrix of a dynamical system. 
We show how to obtain the gradient of the Lagrangian functional of a generic optimal control problem using the adjoint method, where the control variables are multiplicative inputs. 
We illustrate the theory with the precision cost that aims at steering the nodal states to prescribed target trajectories. 
We first focus on a linear time-variant system for which further expressions of the gradient can be obtained. 
Second, we uncover the temporal interaction that pushes a nonlinear neural units model towards target trajectories.  
For both systems, we numerically solve the optimal control problem, imposing various constraint on the structure of the controlled edges. 
The workflow presented here can be leveraged to steer the dynamics of systems with artificial or engineered interaction that is tunable.   
\end{abstract} 

\maketitle

\section{Introduction}
The collective behavior displayed by coupled dynamical systems is the result of the interplay between internal parameters of the individual components and the interaction among them~\cite{pikovsky2001synchronization,newman2018networks}. 
Using complex networks, individual units are typically modeled as nodes while the interaction is represented by edges. 
While in some cases or approximations, the interaction can be taken as time-independent, -- the structure and the weight of the edges remain constant over time -- it is not an accurate modelling for systems in various fields~\cite{holme2012temporal}. 
For example, in social interactions, people exchange social cues with different individuals over the course of time~\cite{starnini2025opinion}; or in neuronal networks, where synaptic connectivity evolves according to plasticity rules that depend on the activity of the neurons~\cite{clark2024theory}. 
The interaction can also be modulated in engineered systems where individual units share information, e.g. in vehicular platoons or swarming robots~\cite{vasarhelyi2018optimized}. 
In general, the time-dependence of the interaction enables the system to access states that could not be reached on a static network~\cite{baumann2020periodic}. 
That makes the interaction a highly relevant variable to control. 
Nevertheless, control frameworks proposed so far essentially considered additive node inputs as control variables with a time-varying interaction network~\cite{posfai2014structural,zhang2021higher}. 
Of course, controlling the interaction might be challenging in practice and sometimes not possible. 
For example, the capacity of a transmission line in a power grid cannot be controlled online. 
However, the interaction within swarms of robots and flocks of autonomous vehicles, or social networks interaction between individuals can be controlled in principle. 
Also, advances in neuromorphic devices and optical control of neuronal channels and receptors allow to control the coupling between neurons and tune synaptic dynamics~\cite{kuzum2012nanoelectronic,paoletti2019optical,liu2023artificial}. 

Here, we investigate how one can steer the dynamics of a system to desired states by controlling the interaction network. 
Such network control strategy is typically more challenging than traditional additive control, mostly because of its multiplicative nature. 
Previous works have used a related approach to design networks that realize specific input-output relationships~\cite{farotimi1989neural}, select edge weights that will enable a consensus to be reached in discrete systems~\cite{el2014optimal}, or find adaptive edge weights rules that will enforce synchronization~\cite{yu2012distributed}.
In this work instead, we want to find the time-dependent coupling weights that will steer the node dynamics to prescribed trajectories or enforce global collective features such as synchronization. 
Using a nonlinear optimal control framework and a general form of coupled dynamical system, we define the Lagrangian functional of the control problem that accounts for costs from both performing the desired task, and applying the control input. 
Then, we derive the adjoint state evolution to obtain an expression for the gradient of the total cost, which can be used to design control strategies for generic optimal control problems. 
To apply the method on concrete examples, we first focus on a linear time-variant dynamics. 
The latter, together with their time-invariant counter part are used in various fields such as opinion dynamics~\cite{baumann2020laplacian, starnini2025opinion}\,, network diffusion~\cite{tyloo2022layered}\,, system control~\cite{marino2003adaptive}\,.
For this system, we derive the gradient of the Lagrangian functional including the precision cost and investigate how the states of such networked dynamics can be steered to desired trajectories by controlling the entire interaction, or only a fraction of it. 
Second, we apply the framework to a nonlinear model of neural units that has been used to describe the activity of firing neurons~\cite{sompolinsky1988chaos} and in other interdisciplinary contexts~\cite{starnini2025opinion}. 
Similarly, for this nonlinear system, we calculate the gradient of the Lagrangian functional including the precision cost and numerically test the interaction control under various constraints on the number of controlled matrix elements. 

The manuscript is organized as follows. In Sec.~\ref{sec1}, we first define a general form of coupled dynamical systems and present the optimal control framework using the adjoint method. 
Then, in the same section, we sequentially introduce the linear time-variant dynamics and neural units system, and apply the framework to control these systems. 
In Sec.~\ref{sec:num}, we numerically test the proposed framework and explore different constrain on the controlled interaction network. 
Finally, we discuss the results and the outlook in Sec.~\ref{sec:dis}\,.


\section{Methods}\label{sec1}
In this section, we introduce a framework for controlling the state of a networked system by controlling its interaction. 
First, we consider a general form of nodal dynamics and performance metrics and derive the gradient of the Lagrangian functional. 
Then, we apply the formalism to a linear time-variant system and a neural units system, in order to steer their dynamics to prescribed target trajectories.
\subsection{Network controlled system}
We consider a networked system composed of $N$ nodes, each of them with a state denoted ${\bf x}_i\in\mathbb{R}^M$\,. 
The coupling between the nodes is given by a matrix of time-dependent components $W_{ij}(t)\in \mathbb{R}$, $i,j=1,...N$\,. 
For now, we do not impose any conditions on the diagonal of ${\bf W}(t)$\,, which impacts the self-interaction of the nodes. 
But in the next sections, we will impose that the diagonal vanishes so that the control is applied only on the interaction between nodes, and not on their internal dynamics. 
Altogether, the dynamics of the networked system is given by,
\begin{align}\label{eq:main1}
    \dot{\bf x}_i = {\bm F}_i({\bf x}_i) +  \sum_{j=1}^N W_{ij}(t) {\bm H}_i({\bf x}_i, {\bf x}_j)\,, i = 1,...N\,,
\end{align}
where ${\bm H}_i({\bf x_i}, {\bf x_j})$ is the coupling function between the states of node $i$ and $j$\,. 
The internal node dynamics is modelled by the dynamical flows ${\bm F}_i({\bf x}_i)$\,. 
Both ${\bm H}_i$'s and ${\bm F}_i$'s are differentiable function of their arguments. 
We also assume that we have a fixed initial condition ${\bf x}(t=0)={\bf x}^{(0)}$\,. 
As we are interested in controlling the state by acting on the connectivity, there is no additive control term in Eq.~(\ref{eq:main1})\,. 
Instead, we assume that $W_{ij}(t) = U_{ij}(t)$ for $i,j=1,...,N$\,, i.e. both the self-interaction and the coupling between the nodes are controlled. 
In the next sections, we will explore more restrictive scenarios where only parts of the ${\bf W}(t)$ are controlled. 
In the following, we equally denote the coupling within the nodes ${\bf U}$ or ${\bf W}$\,. 
Let us now present the framework for nonlinear optimal control based on the controlled interaction.

\subsection{Nonlinear network optimal control framework}\label{Sec:2}
The first step to design an optimal control signal is to define a performance metric that quantifies optimality. 
In general, the performance metric is given by two terms: one describing the control effort, that we denote $\mathcal{C}_u$\,, and one that evaluates how well a specific task is achieved, which we denote $\mathcal{C}_e$\,. 
Eventually, one has the total performance metric,
\begin{align}\label{eq:PM}
    \mathcal{C}({\bf x}\,,{\bf U}) = \mathcal{C}_u({\bf U}) + \mathcal{C}_e({\bf x})\,,
\end{align}
where one can further define $\mathcal{C}_u({\bf U}) = \int_{t_u}^T c_u(t){\rm d}t$\,, $\mathcal{C}_e({\bf x}) = \int_{t_e}^T c_e(t){\rm d}t$\,, with $t_u \le t_e$\,, respectively the times when the control and when the performance start to be measured. 
This means that the interval over which the control is applied does not need to be the same as the interval when we assess the performance metric. 
At this stage, it is not necessary to explicitly define $\mathcal{C}$\,. 
We rather keep the formalism general and pick specific functionals in later sections. 

Second, to obtain the optimal control signal ${\bf U}(t)$\,, one defines the Lagrangian functional of the controlled system which reads,
\begin{align}\label{eq:lag}
    \mathcal{L}({\bf x}, \dot{\bf x}, {\bf U}) &= \int_0^T {\bf 1}_{[t_u,T]}c_u(t) + {\bf 1}_{[t_e,T]}c_e(t) \\
    &+ {\bm \lambda}^\top {\bf g}({\bf x}, \dot{\bf x}, {\bf U}) \,{\rm d}t'\,,\nonumber
\end{align}
where ${\bf g}_i({\bf x}, \dot{\bf x}, {\bf U}) = \dot{\bf x}_i - {\bm F}_i({\bf x}_i) -  \sum_{j=1}^N U_{ij}(t) {\bm H}_i({\bf x}_i, {\bf x}_j)$ for $i=1,...N$\,, such that the overall length of ${\bf g}$ is $NM$\,. 
The indicator function is denoted ${\bf 1}_{[.,.]}$\,. 
The adjoint state (Lagrange multipliers) is denoted ${\bm \lambda}(t) = [{\underline{\bm \lambda_1}}, ..., {\underline{\bm \lambda_N}}]\in \mathbf{R}^{NM}$\,. 
The functional derivative of the Lagrangian with respect to the $U_{ij}$ is given by,
\begin{align}\label{eq:grad1}
    \frac{{\delta} \mathcal{L}}{{\delta}U_{ij}} =  {\bf 1}_{[t_u,T]}\frac{{\partial c_u}}{\partial U_{ij}} - \underline{\bm \lambda}_i^\top {\bm H}_i({\bf x}_i, {\bf x}_j)\,,  
\end{align}
where the adjoint state evolution is obtained from the Euler-Lagranges equations as,
\begin{align}\label{eq:adj}
    \dot{\bm\lambda}^\top \frac{\partial{\bf g}}{\partial \dot{\bf x}} = {\bf p}({\bf x}) + {\bm \lambda}^\top \frac{\partial{\bf g}}{\partial {\bf x}} - {\bm \lambda}^\top\frac{{\rm d}}{{\rm d}t}\frac{\partial{\bf g}}{\partial \dot{\bf x}}\,,
\end{align}
with ${\bf p}({\bf x}, \dot{\bf x})$ defined such that
\begin{align}\label{eq:gru}
    \frac{{\delta} \mathcal{C}_e}{{\delta}U_{ij}}= {\bf 1}_{[t_e,T]}\,{\bf p}({\bf x}) \frac{{\rm d}{\bf x}}{{\rm d}{U_{ij}}}  \,,
\end{align} 
namely, ${\bf p} = \frac{\partial c_e}{\partial {\bf x}}$\,.  
The terminal condition for the adjoint state is obtained from the optimality condition,
\begin{align}
    \frac{{\rm d}}{{\rm d}\epsilon}&\mathcal{L}({\bf x}+\epsilon{\bm\delta}, \dot{\bf x}+\epsilon\dot{\bm\delta}, {\bf U})|_{\epsilon=0} \\
    &  =\int_0^T \left[{\bf p}({\bf x}) + {\bm \lambda}^\top \frac{\partial{\bf g}}{\partial {\bf x}}\right]{\bm\delta} + {\bm \lambda}^\top\frac{\partial{\bf g}}{\partial \dot{\bf x}}\dot{\bm\delta}\,{\rm d}t'  \nonumber\\
    &  =\int_0^T \left[{\bf p}({\bf x}) + {\bm \lambda}^\top \frac{\partial{\bf g}}{\partial {\bf x}}- \frac{{\rm}d}{{\rm d}t}{\bm \lambda}^\top\frac{\partial{\bf g}}{\partial \dot{\bf x}}\right]{\bm\delta}\,{\rm d}t' \nonumber\\
    &+ {\bm \lambda}^\top(T)\frac{\partial{\bf g}}{\partial \dot{\bf x}}{\bm\delta}(T) - {\bm \lambda}^\top(0)\frac{\partial{\bf g}}{\partial \dot{\bf x}}{\bm\delta}(0)  \nonumber\\
    &= {\bm \lambda}^\top(T)\frac{\partial{\bf g}}{\partial \dot{\bf x}}{\bm\delta}(T) = 0\,, \nonumber
\end{align}
where $\epsilon$ is a scalar, $\bm\delta$ is smooth perturbation that has to satisfy the initial condition, i.e. ${\bm\delta}(0)=0$\,. 
In the second equality above, we used the evolution of the adjoint given by Eq.~(\ref{eq:adj})\,.
Because ${\bm\delta}(T)$ is free, the optimality condition imposes the terminal condition for the adjoint state ${\bm \lambda}(T) = 0$\,.
Taking into account the expression for $\bf g$\,, the evolution of the adjoint state Eq.~(\ref{eq:adj}) simplifies to,
\begin{align}\label{eq:adj_s1}
    \dot{{\bm \lambda}}^\top  = {\bf p}({\bf x}) + {\bm\lambda}^\top \frac{\partial{\bf g}}{\partial {\bf x}} \,.
\end{align}
The latter equation can be solved backward in time, starting from the terminal condition ${\bm \lambda}(T)=0$ to obtain the trajectory of the adjoint state. 
Then, it can be injected into Eq.~(\ref{eq:grad1}) to obtain the gradient of the Lagrangian functional with respect to $\bf U$\,. 
Note that one does not need to evaluate Eq.~(\ref{eq:gru})\,, but only need to find $\bf p$\,. 

In summary, to obtain the optimal control in terms of the performance metric Eq.~(\ref{eq:PM})\,, one sequentially performs the following steps:
\begin{enumerate}
    \item Solve the state dynamics of the system Eq.~(\ref{eq:main1}) forward from the initial condition ${\bf x}(t=0)={\bf x}^{(0)}$\,;
    \item Use the forward state solution to solve the evolution of the adjoint state Eq.~(\ref{eq:adj_s1}) backward in time starting from the terminal condition ${\bm \lambda}(T)=0$\,;
    \item Compute the gradient Eq.~(\ref{eq:grad1}) using the adjoint state and update the control signal ${\bf U}(t)$\,;
    \item Go back to step 1 and repeat until convergence.
\end{enumerate}
This procedure is usually called the \textit{adjoint method}~\cite{belegundu1985lagrangian,giles2000introduction}\,, and has been widely used in recent applications such as machine learning. 
See Ref.~\cite{salfenmoser2024framework} for an overview of its application for additive control inputs in neural population models. 

Many different algorithms exist to perform the optimization of the performance metric based on the gradient. 
Also, because the problem in typically non-convex, it is likely that one obtains a local minimum of the performance metric instead of the global one (See Sec.~\ref{sec:lo} below). 
The framework presented so far is general for a dynamical system of the form given in Eq.(\ref{eq:main1}) and rather general cost functionals. 
Below we focus first on a linear time-variant dynamics and then on a neural units system. 
The costs we are going to investigate are specified below as well.
Note that one could include a terminal cost on the state of the system by considering the Lagrangian functional
\begin{align}
    \mathcal{L}({\bf x}, \dot{\bf x}, {\bf U}) &= \Phi({\bf x}(T)) + \int_0^T {\bf 1}_{[t_u,T]}c_u(t) + {\bf 1}_{[t_e,T]}c_e(t) \nonumber \\
    &+ {\bm \lambda}^\top {\bf g}({\bf x}, \dot{\bf x}, {\bf U}) \,{\rm d}t'\,, \nonumber
\end{align}
where $\Phi({\bf x}(T))$ is the terminal cost. 
In the formalism described above, such terminal cost affects the terminal condition of the adjoint state and imposes ${\bm \lambda}(T) = \frac{\partial \Phi}{\partial {\bf x}}({\bf x}(T))$\,.

\subsection{Network controlled linear time variant dynamics}\label{sec:li}
We consider a specific linear time-variant dynamics that fixes the internal dynamics and coupling function in Eq.(\ref{eq:main1})\,. 
The networked system is composed of $N$ nodes, each of them with a scalar state denoted ${ x}_i\in\mathbb{R}$\,. 
The coupling between the nodes is given a matrix of time-dependent components $W_{ij}(t)\in \mathbb{R}$, $i,j=1,...N$\,. 
The dynamics of the system is given by,
\begin{align}\label{eq:main}
    \dot{ x}_i =  \alpha\, (x_i^* - x_i) +  \sum_{j=1}^N W_{ij}(t) {x}_j\,, i = 1,...N\,,
\end{align}
where $x_i^*\in\mathbb{R}$ is the natural state towards which the $i$-th node is brought back to at a rate $\alpha$ when there is no coupling to other nodes. 

The Lagrangian functional for the linear time-variant dynamics reads,
\begin{align}\label{eq:lag1}
    \mathcal{L}({\bf x}, \dot{\bf x}, {\bf U}) &= \int_0^T {\bf 1}_{[t_u,T]}c_u(t) + {\bf 1}_{[t_e,T]}c_e(t) \\
    &+ {\bm \lambda}^\top {\bf g}({\bf x}, \dot{\bf x}, {\bf U}) \,{\rm d}t'\,,\nonumber
\end{align}
where ${g}_i({\bf x}, \dot{\bf x}, {\bf U}) = \dot{ x}_i - \alpha (x_i^* - { x}_i) -  \sum_{j=1}^N U_{ij}(t) {x}_j$ for $i=1,...N$\,.
Here, the adjoint state ${\bm \lambda}(t)\in \mathbb{R}^{N}$\,. 
The gradient of the Lagrangian functional with respect to the $U_{ij}$ is given by,
\begin{align}\label{eq:grad}
    \frac{{\delta} \mathcal{L}}{{\delta}U_{ij}} =  {\bf 1}_{[t_u,T]}\frac{{\partial c_u}}{\partial U_{ij}} - { \lambda}_i {x}_j\,.
\end{align}
The theory presented in the previous section remained general in terms of dynamical system and performance metric for the control. 
Here, we want to narrow down the analysis to enforce the node variables to follow specific trajectories by controlling the network edges.  
To do that, we define the performance metric over the node state variables ${\bf x}$\,,
\begin{align}\label{eq:ce}
    {\mathcal{C}}_e &= \frac{w_e}{2N}\int_{t_e}^T\sum_{j=1}^N (x_j - \tilde{x}_j)^2\, {\rm d}t\,,
\end{align}
where the sum runs over all the network nodes, we introduced the weight $w_e>0$\,, and ${\bf\tilde{x}}(t)$ is the target trajectory. 
The functional Eq.~(\ref{eq:ce}) is usually called \textit{precision cost}.
By minimizing ${\mathcal{C}}_e$\,, one enforces a node activity pattern ${\bf\tilde{x}}(t)$ for $t_e\le t\le T$\,. 
One can further derive the function ${\bf p}({\bf x})$ as,
\begin{align}\label{eq:PM_g}
    {\bf p} = \frac{\delta{{C}}_e}{\delta {\bf x}}  = \frac{w_e}{N}{\bf 1}_{[t_e,T]}({\bf x} - {\tilde{\bf x}})^\top\,.
\end{align}
Typically, one would like to perform this task using as little control input as possible. 
Therefore, for the control cost, we use,
\begin{align}\label{costc}
    {\mathcal{C}}_u &= \frac{w_c}{2N^2}\int_{t_u}^T\sum_{i,j=1}^N U_{ij}^2(t)\, {\rm d}t\,,
\end{align}
where we introduced $w_c>0$ to weight the control cost.
This cost on the control signals affects all the entries of $\bf U$\,, i.e. both the coupling between nodes (off-diagonal terms) and the self-interaction (diagonal terms). 
One can impose more restrictive condition on the control signals. 
For example, the self-interaction can be removed from the control by imposing $U_{jj}=0$ for $j=1,...N$\,. 
Also, one can choose more complicated cost that aims for specific connections between nodes, or even favors sparsity. 
Using the framework presented above one has,
\begin{align}
    \frac{\partial g_i}{\partial x_j}  =  \alpha\, \delta_{ij} - U_{ij}(t)\,,
\end{align}
which allows to derive the adjoint state evolution,
\begin{align}\label{eq:adjT}
    \dot{{\bm \lambda}}  =  \frac{w_e}{N}({\bf x} - {\tilde{\bf x}}) + {\bm \Gamma}(t)^\top{\bm\lambda}  \,.
\end{align}
where ${\bm \Gamma}(t) = \alpha\,\mathbb{I} - {\bf U}(t)$ and $\mathbb{I}$ is the identity matrix.
For this choice of linear time-variant dynamics, one can obtain further expressions for the gradient. 



Indeed, one can conveniently rewrite the overall dynamics Eq.~(\ref{eq:main}) using a vector-matrix form such that the forward dynamics of the system reads,
\begin{align}
\dot{{\bf x}} = \alpha\, ({\bf x}^* - {\bf x}) + {\bf U}\, {\bf x}\,.  
\end{align}
With an initial condition, one can then write down an explicit expression for the states as,
\begin{align}
    {\bf x}(t) = e^{\int_0^t{\bm\Gamma}(t')\, {\rm d}t' }{\bf x}^{(0)} + e^{\int_0^t{\bm \Gamma}(t')\, {\rm d}t' }\int_0^t e^{-\int_0^{t'}{\bm \Gamma}(s)\, {\rm d}s }\alpha{\bf x^*}{\rm d}t'\,,
\end{align}
where ${\bf x}^{(0)}$ is the initial state.

Then, using Eq. (\ref{eq:adjT}) one can write an explicit expression for the adjoint state as,
\begin{align}
    {\bm \lambda}(t) &= e^{\int_0^t{\bm \Gamma}(t')^\top\, {\rm d}t' }{\bm \lambda}^{(0)} \\
    &+ e^{\int_0^t{\bm \Gamma}(t')^\top\, {\rm d}t' }\int_0^t e^{-\int_0^{t'}{\bm \Gamma}(s)^\top\, {\rm d}s }[{\bf x}(t') - {\tilde{\bf x}}(t')]{\rm d}t'\,, \nonumber
\end{align}
where the initial condition is obtained by imposing the terminal condition ${\bm \lambda}(T) = 0$ such that ${\bm \lambda}^{(0)} = -\int_0^T e^{-\int_0^{t'}{\bm \Gamma}(s)^\top\, {\rm d}s }[{\bf x}(t') - {\tilde{\bf x}}(t')]{\rm d}t'$\,.
Now that we have expressions for both the state and the adjoint state of the system, we can inject these solutions into the gradient of the Lagrangian functional Eq.~(\ref{eq:grad}) such that,
\begin{align}\label{eq:gra_sm}
     \frac{{\delta} \mathcal{L}}{{\delta}U_{ij}} = \frac{w_c}{N^2}{\bf 1}_{[t_u,T]}U_{ij} - {\lambda}_i x_j \,,\, i,j=1,...N\,.
\end{align}
Starting from an initial guess for the control signal $\bf U$\,, one can numerically compute Eq.~(\ref{eq:gra_sm})\, and use it to find a minimum of the performance metric. 
We numerically illustrate different control signals in Sec.~{\ref{sec:num}}\,, where we explore different constraints on the network connectivity. 


\begin{figure}
    \centering
    \includegraphics[width=0.99\linewidth]{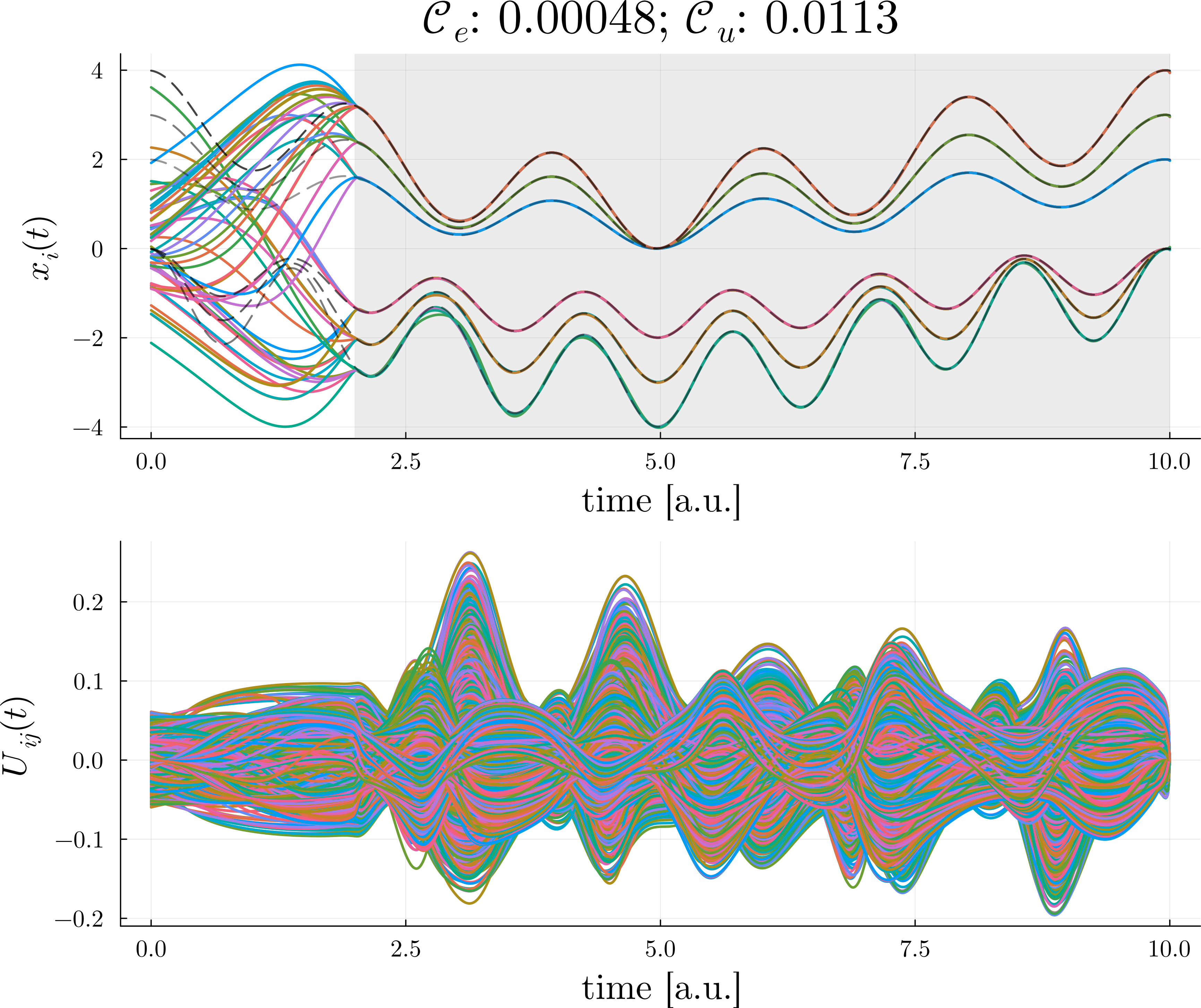}
    \caption{\textbf{Optimal control of the linear time-variant system with all edges controlled}. 
    The top panel shows the state evolution of the $N=50$ nodes (solid lines) and the target signal (dashed grey lines). 
    The bottom panel shows the optimal temporal network found by gradient descent. 
    Only the off-diagonal terms are shown. 
    The gray area shows the time-interval where the precision cost is evaluated. 
    The parameters used for the simulations are given in the main text. 
    Values of the precision cost ($\mathcal{C}_e$) and control cost ($\mathcal{C}_u$) are given above the top panel.}
    \label{fig1}
\end{figure}
\begin{figure*}
    \centering
    \includegraphics[width=0.95\linewidth]{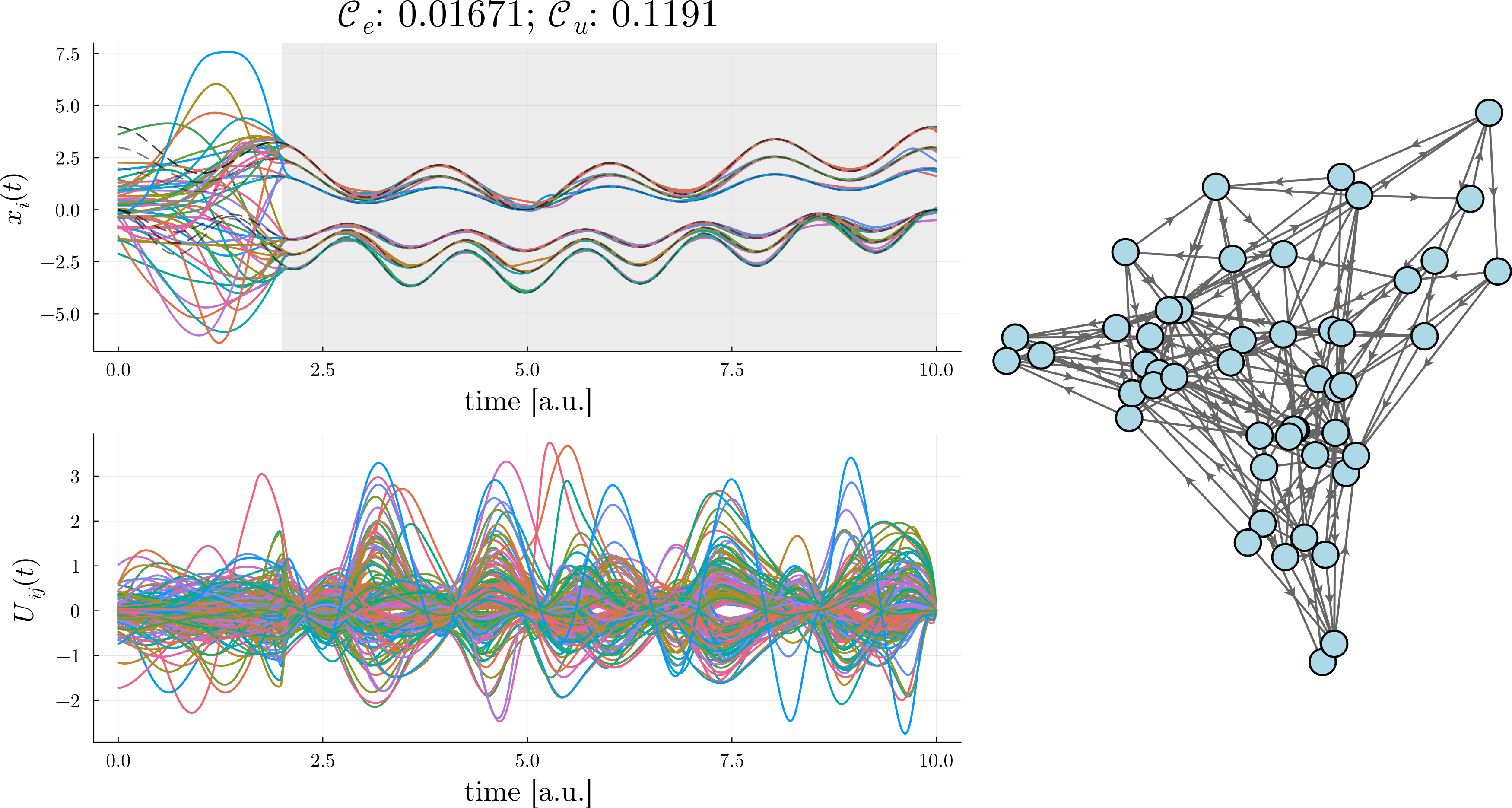}
    \caption{{\textbf{Optimal control of the linear time-variant system with randomly selected controlled edges}}. 
    The top panel shows the state evolution of the $N=50$ nodes (solid lines) and the target signal (dashed grey lines). 
    The bottom panel shows the optimal temporal network found by gradient descent. 
    Only the off-diagonal terms are shown. 
    The gray area shows the time-interval where the precision cost is evaluated. 
    The parameters used for the simulations are given in the main text. 
    The right panel shows the structure of the controlled edges (in grey with arrows indicating the direction of the edges) connecting the nodes (in light blue).   
    Values of the precision cost ($\mathcal{C}_e$) and control cost ($\mathcal{C}_u$) are given above the top panel.}
    \label{fig2}
\end{figure*}
\subsection{Network control of nonlinear neural units}
We now move to a more challenging system to control. 
Indeed, instead of taking a linear time-invariant dynamics, we focus now on a model of $N$ nonlinear neural units governed by the coupled ODEs \cite{sompolinsky1988chaos,clark2024theory}
\begin{align}\label{eq:NLd}
    \dot{x}_i = -\,x_i + J_i\,\phi(x_i) + \sum_{j=1}^N W_{ij}(t)\phi(x_j)\,, i = 1,...N\,,
\end{align}
where $x_i$ is related to the membrane voltage potential of neuron $i$ and $\phi(x_i)$ is a nonlinear gain function that we choose to be $\phi(x)=\tanh(x)$\,. 
The parameters $J_i$'s correspond to the self-interaction of the neural units. 
Such model has also been used in opinion dynamics \cite{starnini2025opinion}. 
The time-dependent coupling matrix is controlled such that $W_{ij}(t) = U_{ij}(t)$ for $i,j=1,...N$\,, with $W_{ii}=0$ for $i=1,...N$\,. 
In the next section, we also consider the case where only a subset of the total possible connections, i.e. the matrix element $W_{ij}(t)$\,, are controlled.  
The interaction and self-interaction allow the degrees of freedom to display complex behaviors~\cite{clark2024theory}\,.
The Langrangian functional of the control problem takes the same form as Eq. (\ref{eq:lag1}) but this time with ${g}_i({\bf x}, \dot{\bf x}, {\bf U}) = \dot{ x}_i + { x}_i - J_i\,\phi(x_i) -  \sum_{j=1}^N U_{ij}(t) \phi({x}_j)$ for $i=1,...N$\,. 
Using Eq. (\ref{eq:grad1})\,, one can derive the gradient of the Langrangian functional and obtain,
\begin{align}\label{eq:gradNL}
    \frac{{\delta} \mathcal{L}}{{\delta}U_{ij}} =  {\bf 1}_{[t_u,T]}\frac{{\partial c_u}}{\partial U_{ij}} - { \lambda}_i \phi({x}_j)\,.
\end{align}
For the performance metric, we consider the same control objective as in Sec.~\ref{sec:li}, namely, the precision cost Eq.~(\ref{eq:ce}).
For the control cost, we use the same functional as in the previous section namely Eq. (\ref{costc})\,. 
To obtain the adjoint state evolution, one needs to calculate the derivative of $\bf g$ with respect to $\bf x$,
\begin{align}
    \frac{\partial g_i}{\partial x_j} = \delta_{ij}(1 - J_i\,\phi'(x_i)) - U_{ij}(t)\phi'(x_j)\,,
\end{align}
with $\phi'(x) = 1 - \tanh^2(x)$\,.
 Including the derivative into the dynamics of the adjoint state Eq. (\ref{eq:adj_s1}) yields,
\begin{align}\label{eq:adjT2}
    \dot{{ \lambda}}_i  &=  \frac{w_e}{N}(x_i - \tilde{x}_i) + \lambda_i \nonumber \\
    &- \lambda_i\left(J_i\,(1-\tanh^2(x_i)) + \sum_{j=1}^N U_{ij}(t) (1-\tanh^2(x_j)) \right) \,,
\end{align}
$i = 1,...N$\,, with the terminal condition ${\bm \lambda}(T) = 0$\,. 
Solving the evolution of the adjoint state allows to compute the gradient of the Lagrangian functional,
\begin{align}\label{eq:gra_sm2}
     \frac{{\delta} \mathcal{L}}{{\delta}U_{ij}} = \frac{w_c}{N^2}{\bf 1}_{[t_u,T]}U_{ij} - {\lambda}_i \tanh(x_j) \,,\, i,j=1,...N\,,
\end{align}
where we used the control cost Eq.~(\ref{costc})\,.
In the next section, we illustrate numerically the optimal control problem based on controlled interaction. 

\begin{figure*}
    \centering
    \includegraphics[width=0.95\linewidth]{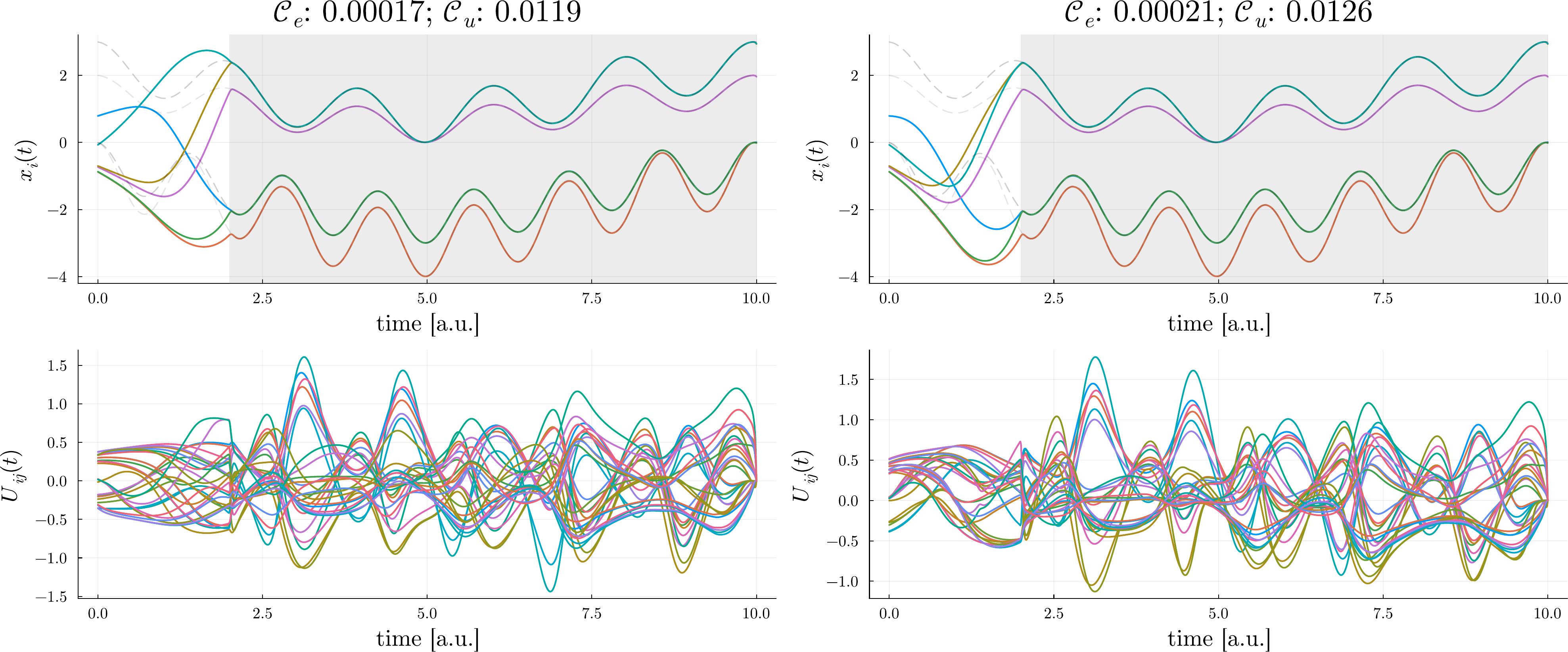}
    \caption{{\textbf{Optimal control of the linear time-variant system for two different initial $U_{ij}$'s}}. 
    The top panel shows the state evolution of the $N=6$ nodes (solid lines) and the target signal (dashed grey lines). 
    The bottom panel shows the optimal temporal network found by gradient descent. 
    Only the off-diagonal terms are shown. 
    The gray area shows the time-interval where the precision cost is evaluated. 
    All the edges are controlled. 
    The parameters used for the simulations are given in the main text.   
    Values of the precision cost ($\mathcal{C}_e$) and control cost ($\mathcal{C}_u$) are given above the top panel.}
    \label{fig3}
\end{figure*}

\begin{figure}
    \centering
    \includegraphics[width=0.99\linewidth]{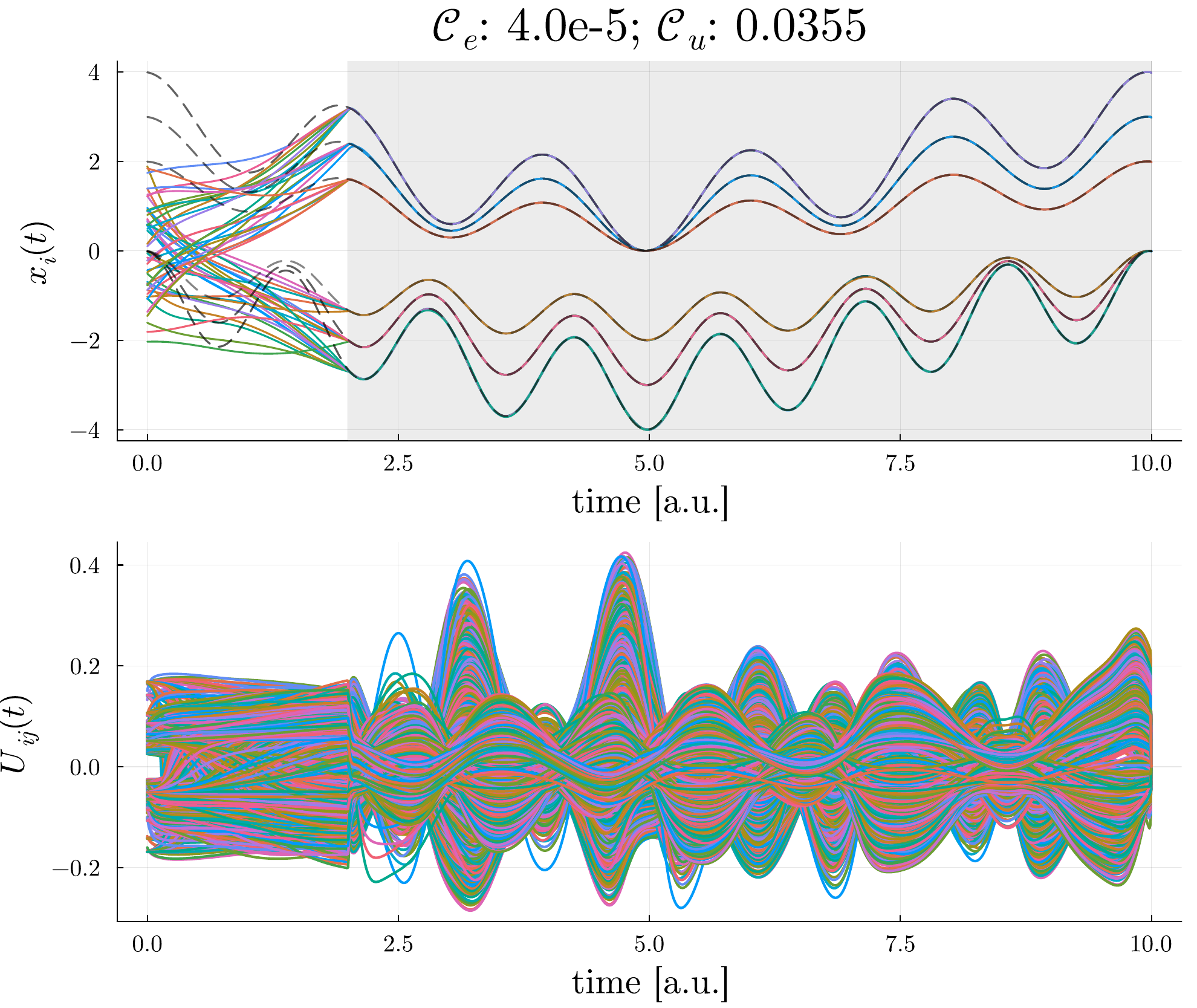}
    \caption{\textbf{Optimal control of the neural units system with all edges controlled}. 
    The top panel shows the state evolution of the $N=50$ nodes (solid lines) and the target signal (dashed grey lines). 
    The bottom panel shows the optimal temporal network found by gradient descent. 
    Only the off-diagonal terms are shown. 
    The gray area shows the time-interval where the precision cost is evaluated. 
    The parameters used for the simulations are given in the main text. 
    Values of the precision cost ($\mathcal{C}_e$) and control cost ($\mathcal{C}_u$) are given above the top panel.}
    \label{fig4}
\end{figure}
\begin{figure}
    \centering
    \includegraphics[width=0.99\linewidth]{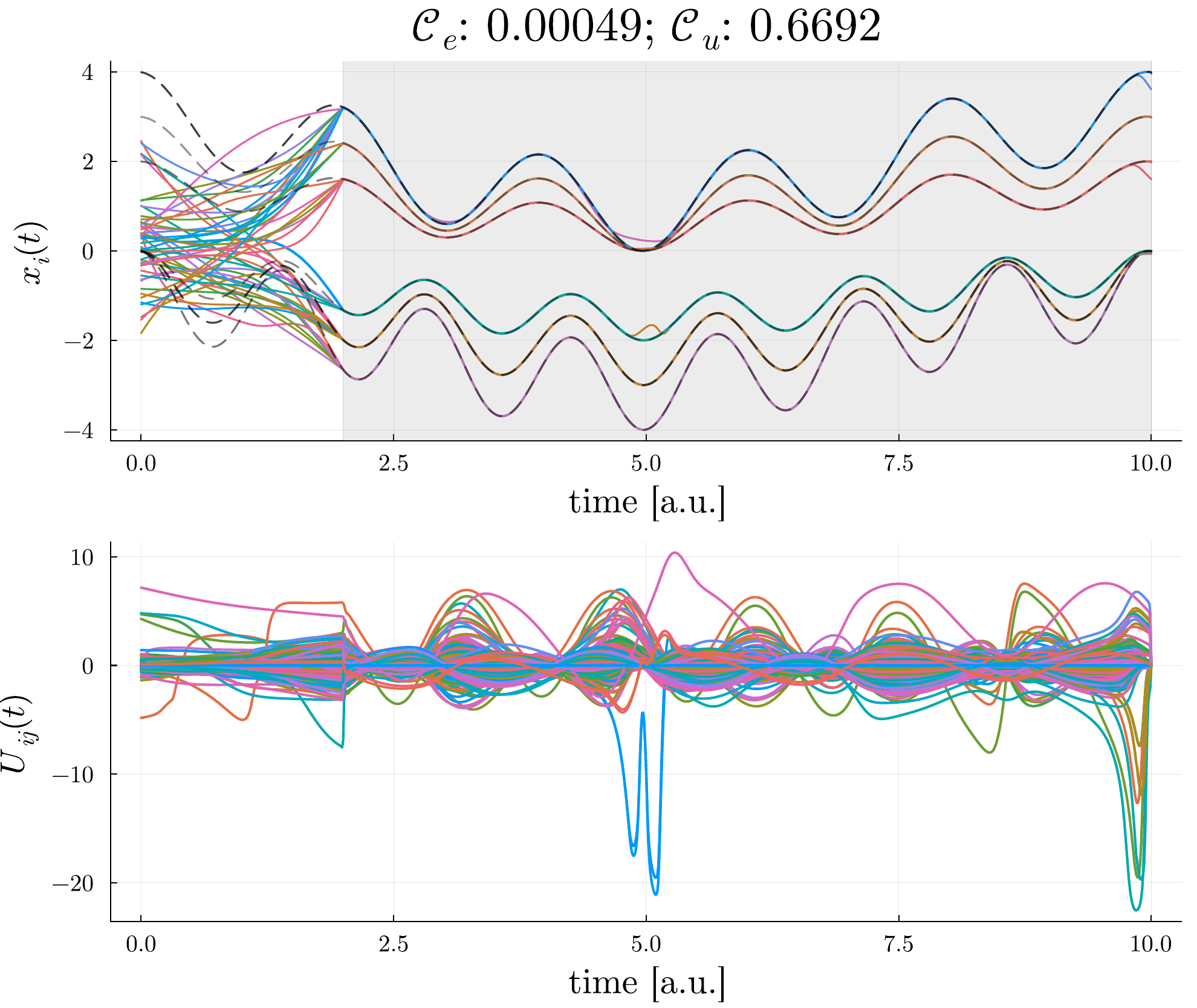}
    \caption{{\textbf{Optimal control of the neural units system with randomly selected controlled edges}}. 
    The top panel shows the state evolution of the $N=50$ nodes (solid lines) and the target signal (dashed grey lines). 
    The bottom panel shows the optimal temporal network found by gradient descent. 
    Only the off-diagonal terms are shown. 
    The gray area shows the time-interval where the precision cost is evaluated. 
    The structure of the controlled edges is shown in the right panel of Fig.~\ref{fig2}\,.
    The parameters used for the simulations are given in the main text. 
    Values of the precision cost ($\mathcal{C}_e$) and control cost ($\mathcal{C}_u$) are given above the top panel. 
    For this simulation, we stopped the gradient descent when it reached 30'000 iterations with $\max_{i,j,t}\left|\frac{\delta\mathcal{L}}{\delta U_{ij}}\right|\cong 3\times 10^{-3}$ }
    \label{fig5}
\end{figure}

 \section{Numerical results}\label{sec:num}
 For all the simulations below, at each iteration, one numerically solves the forward and backward dynamics with the current $U_{ij}(t)$'s. 
 This yields an interpolation of the system's state and the adjoint state over the interval $[0, T]$\,.  
 Then, $U_{ij}(t)$'s and the gradient are numerically obtained at discrete equidistant time points in the interval $[0, T]$. 
 The control for the next iteration is obtained by updating the interaction with the gradient. 
In all the simulations discussed below, the optimal control signals $U_{ij}(t)$ have been obtained by gradient descent using Adam~\cite{kingma2014adam}. 
The initial state of the control variables for the gradient descent is taken as homogeneously vanishing, i.e. $U_{ij}(t)=0$ for $i,j=1,...N$\,, $t\in [0,T]$\,, unless specified. 
We stop the gradient descent when $\max_{i,j,t}\left|\frac{\delta\mathcal{L}}{\delta U_{ij}}\right|<10^{-3}$\,.
 For the cost functional, we use the weights for the control $w_c=5\times10^{-4}$ and the precision cost $w_e=1$ for all simulations. 

 The implementation of the method is available here~\cite{github}.
 \subsection{Linear time-variant system}
 We consider a set of $N=50$ nodes following the dynamics Eq. (\ref{eq:main}) with natural states that are normally distributed around the origin with $x_i^*\sim \mathcal{N}(0,1)$ for $i=1,...N$\,, and the gain set to $\alpha=1$\,. 
The initial condition is simply set to be the natural states, i.e. ${\bf x}^{(0)} = {\bf x}^*$\,.
To illustrate how the framework developed above for the precision cost, we consider a set of target trajectories towards which we aim to steer the linear time-variant system Eq.~\ref{eq:main}\,.
We choose the target signal for the $i$-th node to be 
 \begin{align}\label{tar1}
    \tilde{x}_i(t)\hspace{-1mm} =\hspace{-1mm}
    \begin{cases}
       \hspace{-0.5mm} -1 + [\cos(2\pi 0.3\,t)\cos(2\pi 0.4\,t + 0.1)]\,,\, i=1,...25\,,\\
        \hspace{-0.5mm}1 + [\cos(2\pi 0.2\,t)\cos(2\pi 0.3\,t + 0.1)]\,, \, i=26,...50\,,
    \end{cases}
\end{align} 
as it includes different frequencies. 
Below, we consider different constraints on the number of controlled elements of the coupling matrix and their structure. 
As we are interested in controlling the interaction and not the nodal dynamics, we assume that only the off-diagonal terms of $\bf W$ are controlled, i.e. $W_{ij}=U_{ij}$ for $i\neq j$ and $i,j=1,...N$\,, and that $W_{ii}=0$ for $i=1,...N$\,.

\subsubsection{All-to-all interaction control}
Let us start with the less constrained situation where all the off-diagonal elements of the coupling matrix are controlled, i.e. $W_{ij}(t) = U_{ij}(t)$ with $i,j=1,...N$, $i\neq j$\,. 
Figure \ref{fig1} shows the outcome of the optimal control framework. 
The top panel depicts the optimal trajectories obtained for the target Eq.~(\ref{tar1}). 
One observes that the $x_i$'s (solid line) closely match the target trajectories (dashed lines) during the time interval where the precision cost is measured, i.e. $t_e=2\le t\le T=10$\,. 
This is also confirmed by the almost vanishing value of $\mathcal{C}_e$\,. 
The bottom panel of Fig.~\ref{fig1} gives the time-dependent couplings $U_{ij}(t)$ obtained by gradient descent.

\subsubsection{Network constrained interaction control}\label{sec:sc}
We now move to the more challenging case where not all the off-diagonal elements of the coupling matrix are controlled. 
In this scenario the controlled elements of the coupling matrix are limited to a subset of all possible interactions. 
To obtain the subset of controlled edges used in Fig.~\ref{fig2}\,, we randomly added controlled edges until all the nodes reached a non-vanishing in-degree. 
For the random seed used, the latter happened after selecting 195 controlled edges over the 2450 possible edges (less than 8\% of edges are controlled). 
The corresponding structure is illustrated in the right panel of Fig.~\ref{fig2}\,.
The solution found for the optimal control control problem is given in the left panel of Fig.~\ref{fig2}\,.  
Note that these edges are directed, i.e. if $U_{ij}$ is controlled, it does not mean that $U_{ji}$ is necessary controlled too.
The control objective is achieved with a low value for the precision cost. 
The control cost $\mathcal{C}_u$ for these instances is higher in the network constrained case compared to the all-to-all case (Fig~\ref{fig1})\,. 
In general, this is expected as the number of controlled edges is lower in the constrained case compared to the all-to-all case. 
Thus, to reach the same objective, more effort is required from each edge in Fig.~\ref{fig2}\,.

\subsubsection{Local optimum}\label{sec:lo}
To illustrate the fact that the optimal control problem considered here is non-convex in general, we initiated the gradient with two different state $U_{ij}(t)$'s. 
Figure~\ref{fig3} shows two different local minima obtained at the end of the gradient descent. 
In both cases the target trajectories are well enforced. One observes differences in the optimal control $U_{ij}(t)$'s in the bottom panels.


\subsection{Nonlinear neural units}
We apply interaction control to push $N$ neural units governed by the dynamics Eq.~(\ref{eq:NLd}) to the target trajectories given by Eq.~(\ref{tar1}). 
Similarly to the linear time-variant case, we consider both all-to-all interaction control and then additional constraints on the number of controlled edges. 
The system size is $N=50$ neural units, the self interaction is taken as $J_i\sim \mathcal{N}(0,1)$ for $i=1,...N$\,, and similarly the initial conditions $x^{(0)}_i\sim \mathcal{N}(0,1)$ for $i=1,...N$.
\subsubsection{All-to-all interaction control}
In Fig.~\ref{fig4}\,, we show the outcome of the optimal control problem when all off-diagonal elements of the coupling matrix are controlled and the target trajectories are given by Eq.~(\ref{tar1}). 
Despite the nonlinearity of the neural units dynamics Eq.~(\ref{eq:NLd})\,, one observes that the $x_i$'s follow closely the target. 
This is confirmed by the low value of the precision cost $\mathcal{C}_e$\,.

\subsubsection{Network constrained interaction control}
Let us now move on to the more challenging situation where only a subset of edges are controlled. 
The structural constraint we consider is the same as for the linear time-variant system in Sec.~\ref{sec:sc}\,.
The result of the optimal control problem are shown in Fig.~\ref{fig5}\,. 
One observes that the states of the neural units match well the target trajectories, which is also confirmed by the low values of $\mathcal{C}_e$\,. 
Again, the effort of the control is more important when the structure is constrained compared to the all-to-all case Fig.~\ref{fig1}\,.

 \section{Discussion and outlook}\label{sec:dis}
We have presented a framework for controlling the dynamics of a set of coupled dynamical systems by acting on its interaction. 
Namely, we assume that one has access to the elements of the coupling matrix and leverage them as control variables. 
Using the adjoint method, which is a widely used technique in machine learning, we derived the gradient of the Lagrangian functional of the optimal control problem where the control variables are multiplicative in the dynamics. 
To illustrate the framework, we considered a linear time-variant system and nonlinear coupled neural units. 
For both systems, we pushed the dynamics to target trajectories and found that that one does not need to control all the possible edges to reach the control objective. 
Indeed, we showed that, even when the controlled edges are limited to a subset of the total edges, the objective is still reachable. 
These two examples show the potential of interaction network control to steer complex systems towards target configurations. 
The framework is not limited to the target trajectory Eq.~(\ref{tar1}) and more generally to the precision cost functional considered here. 
Indeed, for any other cost, one needs to find the function $\bf p$ in Eq.~(\ref{eq:gru}) and inject it into the evolution of the adjoint state Eq.~(\ref{eq:adj}). 
By doing so, instead of controlling the trajectories, one could enforce desired global features such as (de-)synchronization. 
Also, by adding a terminal cost to the Lagrangian functional Eq.~(\ref{eq:lag}), one can drive the system to specific target terminal states.  
Moreover, one can consider additional constraints on the problem while performing the gradient descent. 

The current framework could be improved by considering weights for the control cost $w_u$ and the performance metric $w_e$ that evolve with the iterations of the gradient algorithm. 
Indeed, fixing a control cost that is too high sometimes prevents the gradient from reaching a minimum that realizes the objective. 
One could therefore start with a lower control cost and incrementally increase it with the number of gradient steps performed.

Future work should extend the network control framework to include additional constraints coming from specific applications -- in neuroscience there are different ways to emulate or control the connectivity between neurons, each of them coming with their own constraints on $U_{ij}$~\cite{elson1998synchronous,keene2020biohybrid,fu2025constructing}\, -- and also explore more complex systems, both in terms of the their dynamics and their controllable inputs -- combine nodal and interaction control.

\section*{Data Availability Statement}
The code used for the simulations in the Sec.~\ref{sec:num} are available on GitHub~\cite{github}\,. 
No new data were created. 
 
\bibliographystyle{apsrev4-2}
\bibliography{bibliography}

\end{document}